\documentclass[a4paper,11pt]{article}
\usepackage[latin1]{inputenc}
\usepackage[english]{babel}
\usepackage{graphicx}
\usepackage{subfigure}

\begin{document}
\title{{\itshape Free Running Single Photon Detection based on a negative feedback InGaAs APD}}
\author{Tommaso Lunghi$^{a\dagger}$, Claudio Barreiro$^{a}$, Olivier Guinnard$^{a}$,\\ Raphael Houlmann$^{a}$, Xudong Jiang$^{b}$, Mark A. Itzler$^{b}$ and Hugo Zbinden$^{a}$\\\vspace{6pt} $^{a}${\em{Group of Applied Physics, University of Geneva, Geneva CH-1211, Switzerland}};\\ $^{b}${\em{Princeton Lightwave Inc., 2555 US Route 130 S., Cranbury, NJ 08512, USA}}\
\thanks{$^\dagger$ Corresponding author. Telephone: +41223790530. Email:Tommaso.Lunghi@unige.ch \vspace{6pt}}}
\maketitle
\begin{abstract}
InGaAs/InP-based semiconductor avalanche photodiode are usually employed for single-photon counting at telecom wavelength. However they are affected by afterpulsing which limits the diode performance. 
Recently, Princeton Lightwave has commercialised a diode integrating monolithically a feedback resistor. This solution effectively quenches the avalanche and drastically reduces afterpulsing. Here, we report the development and characterization of a detector module based on this diode, implementing an active hold-off circuit which further reduces the afterpulsing and notably improves the detector performances. We demonstrate free-running operation with 600 Hz dark count rate at 10\% detection efficiency. We also improved the standard double-window technique for the afterpulsing characterization. Our algorithm implemented by a FPGA allows to put the APD in a well-defined initial condition and to measure the impact of the higher order afterpulses.\\ 
\end{abstract}

\section{Introduction}

Arguably the best free-running single-photon detectors in the telecom range are superconducting devices, namely the Transition Edge Sensor\cite{Miller}, TES, and the Superconducting Nanowire Single-Photon Detector \cite{Goltsman}, SNSPD. The TESs feature close to unity quantum efficiency and virtually no dark counts, but they are slow.  SNSPDs offer a better trade off between efficiency, dark counts (about 10\% with 10-100 cts/s) maximum count rates (100 MHz) and timing jitter ($<$100 ps). However, due to the need of cryogenic cooling ($<$100 mK and $<$3 K respectively), they are not suitable for most applications. For this reason, InGaAs/InP avalanche detectors driven in Geiger mode (named Single-Photon Avalanche Detector, SPAD) are more frequently used. This detector has the  advantage of a relatively high operation temperature (220-270K with Peltier cooling), reasonably high efficiency (up to 30\%) \cite{Itzler}, high count-rates capability \cite{Jun}, and low timing jitter \cite{Liang}.\\
 Unfortunately these devices suffer from afterpulsing: after the triggering of an avalanche by a photon, another avalanche can be spontaneously generated. This phenomena has been explained with the release of current carriers trapped in the depletion region during the previous avalanches \cite{Kindt}. The number of carriers trapped grows with the avalanche duration, therefore a reduction of the afterpulsing effect is usually achieved by rapidly stopping the current flowing through the SPAD.  This can be most easily done by gating, i.e. applying a bias voltage above breakdown only for a short period. However, this is incompatible with free-running operation, which asks for active or passive quenching \cite{Cova}. While the first technique is difficult to implement if short quenching times are needed ($<$1 ns), the latter is easy but inefficient for standard SPADs. Indeed, discrete components exhibit a large parasitic capacitance, leading leading to excessive charge flow per avalanche \cite{Itzler2}. Only recently has this issue been tackled with the negative feedback avalanche photodiode, NFAD \cite{Itzler2} which consists of a monolithic thin-film resistor integrated directly on the surface of the device. This monolithic integration reduces the parasitic effects, and ideally, the total amount of charges in the diode during the avalanche determined only by the diode depletion capacitance and the excess bias.\\
In this paper, we report the first realization of a detector based on the NFAD by Princeton
Lightwave Incorporation implementing an active hold-off circuit. We perform an exhaustive characterization of the afterpulsing effects using a FPGA-controlled setup. Then we describe a compact free-running module that allows appropriate hold-off time to be applied for optimum performance for a large range of applications.

\section{Negative Feedback Avalanche Photodiode}

\begin{figure}[!htbp]
	\centering
	\subfigure[]{\label{subfig:0:1} \includegraphics[width=5.5cm]{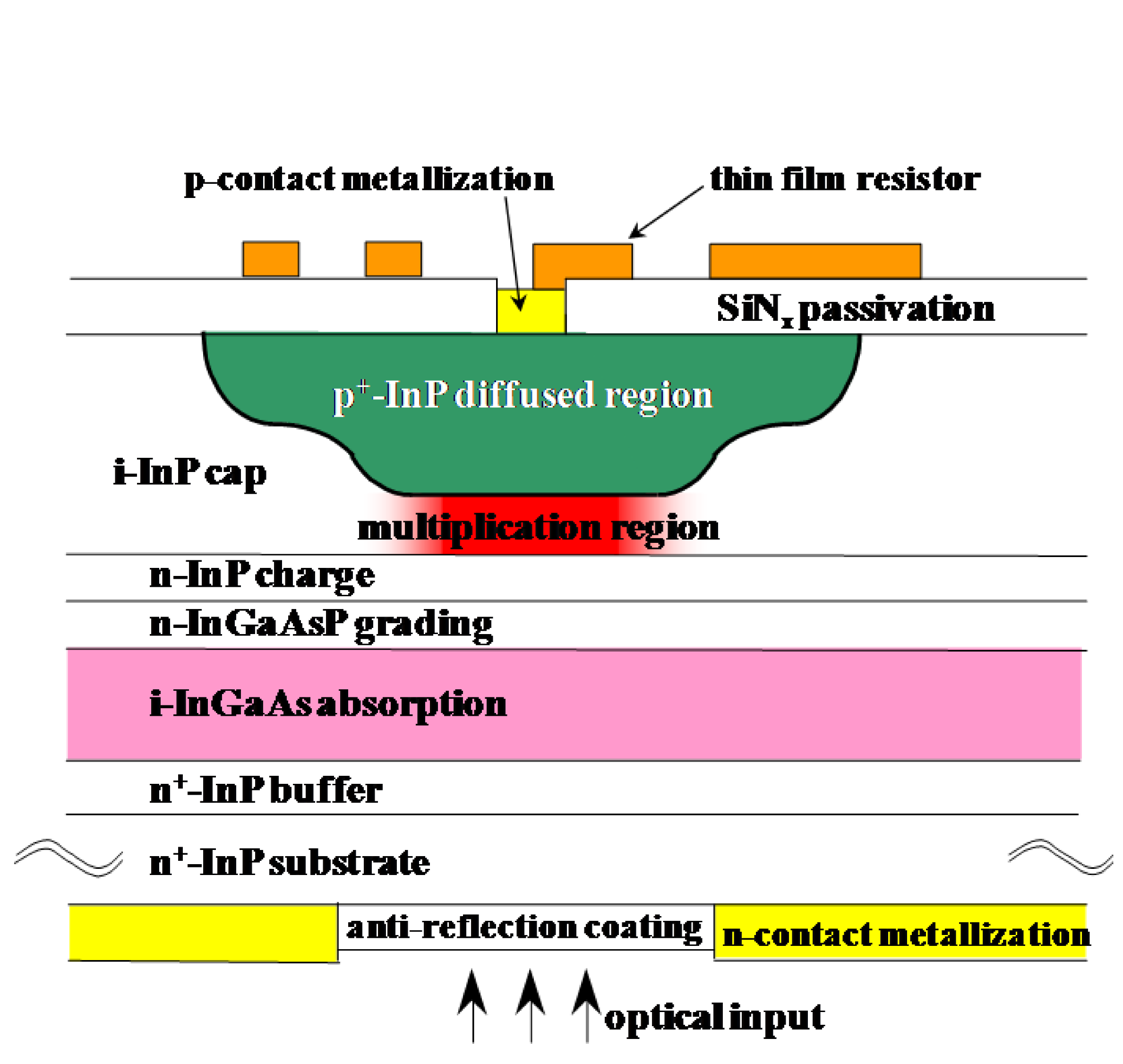}}\qquad
	\subfigure[]{\label{subfig:0:2}\includegraphics[width=5.5cm]{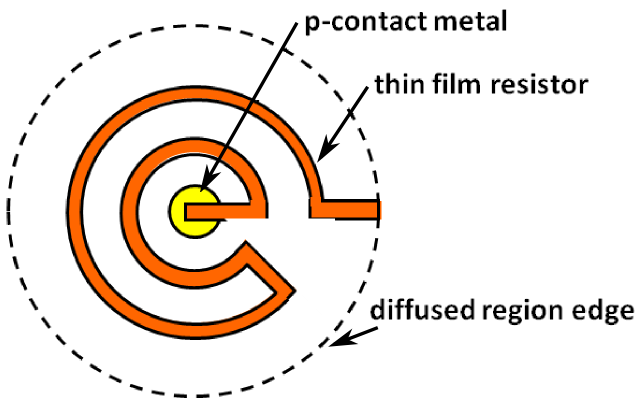}}\\
	\subfigure[]{ \label{subfig:0:3}\includegraphics[width=11 cm]{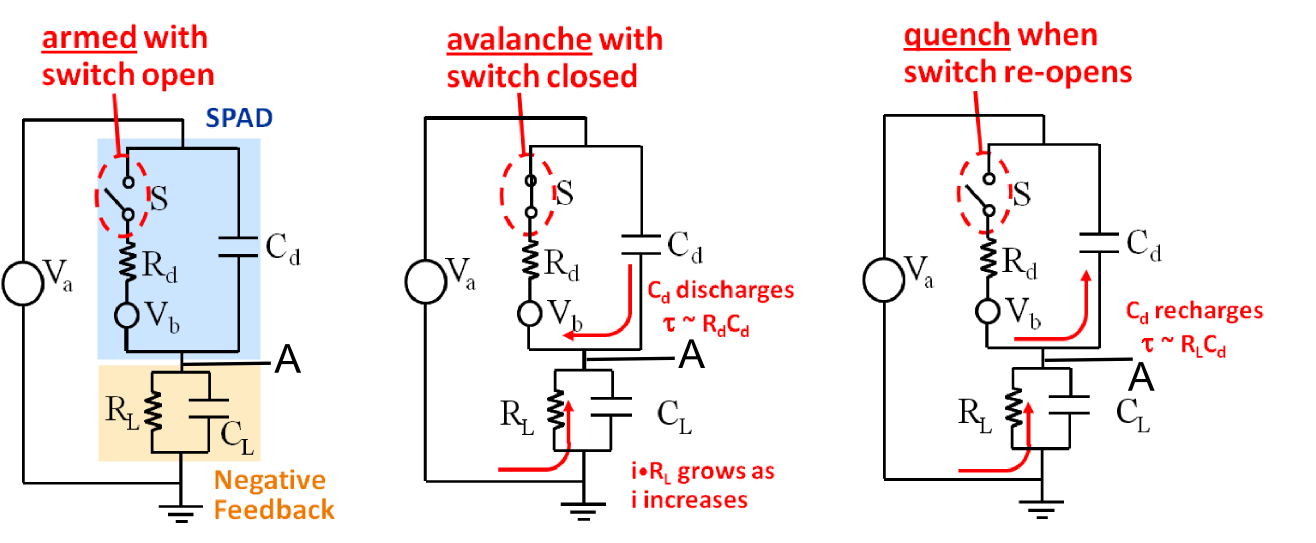}}
	\caption{(a) Cross-sectional schematic representation of the APD growth structure with the monolithically integrated thin-film resistor fabricated on the surface of the device. (b) Top view of the APD device: the orange meander line indicates the geometry of the thin-film resistor. (c) From left to right: equivalent circuit for the negative feedback diode.  This circuit consists of the standard equivalent circuit for the SPAD device (blue background) where R$_d$ and C$_d$ are the equivalent resistance and capacitance characterising the diode. Connected in series to the diode has been displayed (orange background) the negative feedback load with resistor (R$_L$). When the photon triggers the avalanche represented by the closing of the switch S, C$_d$ discharges with characteristic time $\tau_{discharge}$=R$_d$C$_d$ faster than the quench time. The flowing current is quenched by the decreasing of the voltage across the diode yielding V$_a$-IR$_L$. When this voltage is sufficiently close to the breakdown value, S is opened again and the capacitance C$_d$  is recharged with time constant  $\tau_{charge}$=R$_L$C$_d$.}.
	\label{fig:0}
\end{figure}

The key component of the developed module is the NFAD. Figure \ref{subfig:0:1} illustrates the cross-sectional detail of the NFAD.  The NFAD diode epitaxial structure was designed following \cite{Jiang}, and the negative feedback element is a thin-film resistor integrated monolithically with connection to the p-metal contact as shown schematically in Fig.\ref{subfig:0:2}. The chosen resistor length and geometry is a trade off between reducing the quenching time and increasing the recharging time. Figure \ref{subfig:0:3} shows an equivalent circuit of the diode during operating. When an avalanche occurs, the switch S closes and the capacitor associated to the diode (C$_d$) is discharged. The current flow induces a voltage drop across the load resistor R$_L$ decreasing the bias voltage across the diode sufficiently close to the breakdown voltage to allow spontaneous quenching of the avalanche. Increasing R$_L$ reduces the quenching time, however, the re-arming time required to recover the operating condition increases with the resistance, limiting the maximum count rate.\\

The devices under test were PNA-20X Negative Feedback Avalanche Diodes (NFADs) manufactured by Princeton Lightwave. The performance of these devices has been reported recently \cite{Itzler3, Jiang2}. The thin film resistor, $R_{L}$ yields $\sim$950 k$\Omega$ providing rapid quenching time (lower than 1 ns) and long recharging time ($\sim$140 ns).\\
\newline
\section{Afterpulse characterization}

\begin{figure}[!htbp]
	\centering
	\includegraphics[width=4cm]{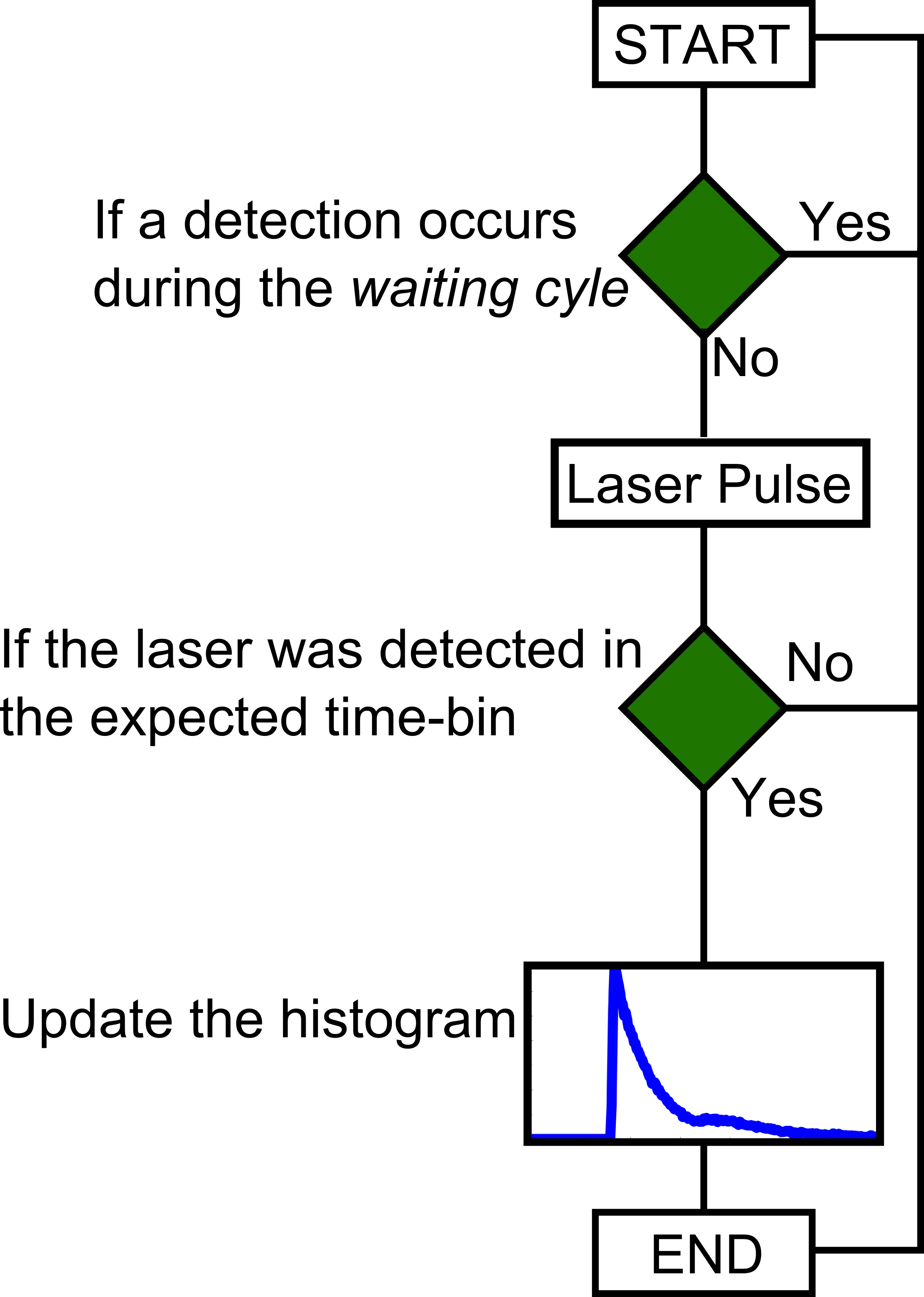}
	\caption{Flow chart of the afterpulse probability measurement procedure}
	\label{fig:10}
\end{figure}

The efficient quenching based on passive monolithically-integrated component leads to small avalanches (10$^5$-10$^6$ carriers \cite{Itzler2}). This is expected to reduce the afterpulsing relative to SPADs operated with greater charge flow per avalanche. In order to verify the efficiency of this technique, we characterized the temporal evolution of afterpulsing as a function of the bias voltage using an improved measurement procedure. Commonly, afterpulsing characterizations are performed by means of the double-window technique \cite{Cova2} which provides the decay curve of minority carriers trapped in the deep levels. However, this procedure cannot characterize higher orders of afterpulses, i.e. releasing  of carriers trapped during an avalanche generated by an afterpulse. These higher order phenomena can significantly change the afterpulsing behaviour and the performance of the detector.\\
In order to characterize this phenomenon, we implement a method designed for free-running, passively quenched APD's. It is controlled by a field programmable gate array integrated circuit, FPGA with clock frequency of 50 MHz. Figure \ref{fig:10} shows the flow chart of the measurement procedure. First, we want to start the measurement with a well-defined initial state where no carriers originating from a former avalanche remain trapped. This is achieved by introducing a preliminary step, named \emph{waiting cycle}, where the FPGA waits until no avalanche has been detected for a given time (defined by the user). If this condition is satisfied the FPGA triggers the laser, otherwise the algorithm is restarted. After sending the laser pulse, the FPGA checks if it has been detected. In this case, after a given hold-off time the bias voltage is increased again and detections occurring during the following 250 time bins are recorded. The time-bin is a multiple of the inverse clock frequency (in our case 20 ns). This means that the acquisition range can be increased at the expense of reduced resolution. The user can choose to include higher order phenomena by recording all the afterpulses occurring after the laser detection. Otherwise only the first event is recorded.\\

\begin{figure}[!htbp]
	\centering
	\subfigure[]{\label{subfig:3:1} \includegraphics[width=7.5cm]{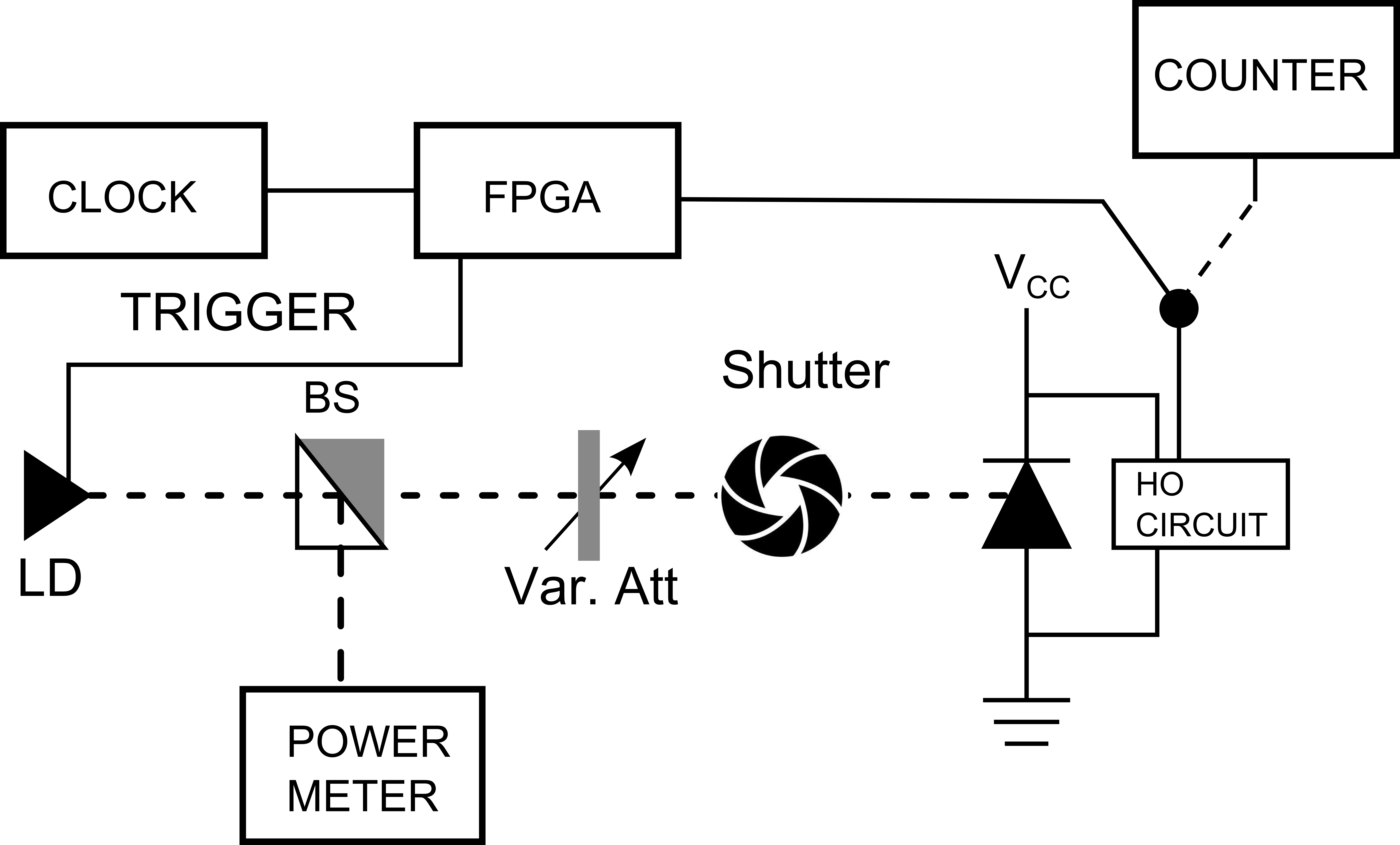}}\qquad
	\subfigure[]{\label{subfig:3:2}\includegraphics[width=7.5cm]{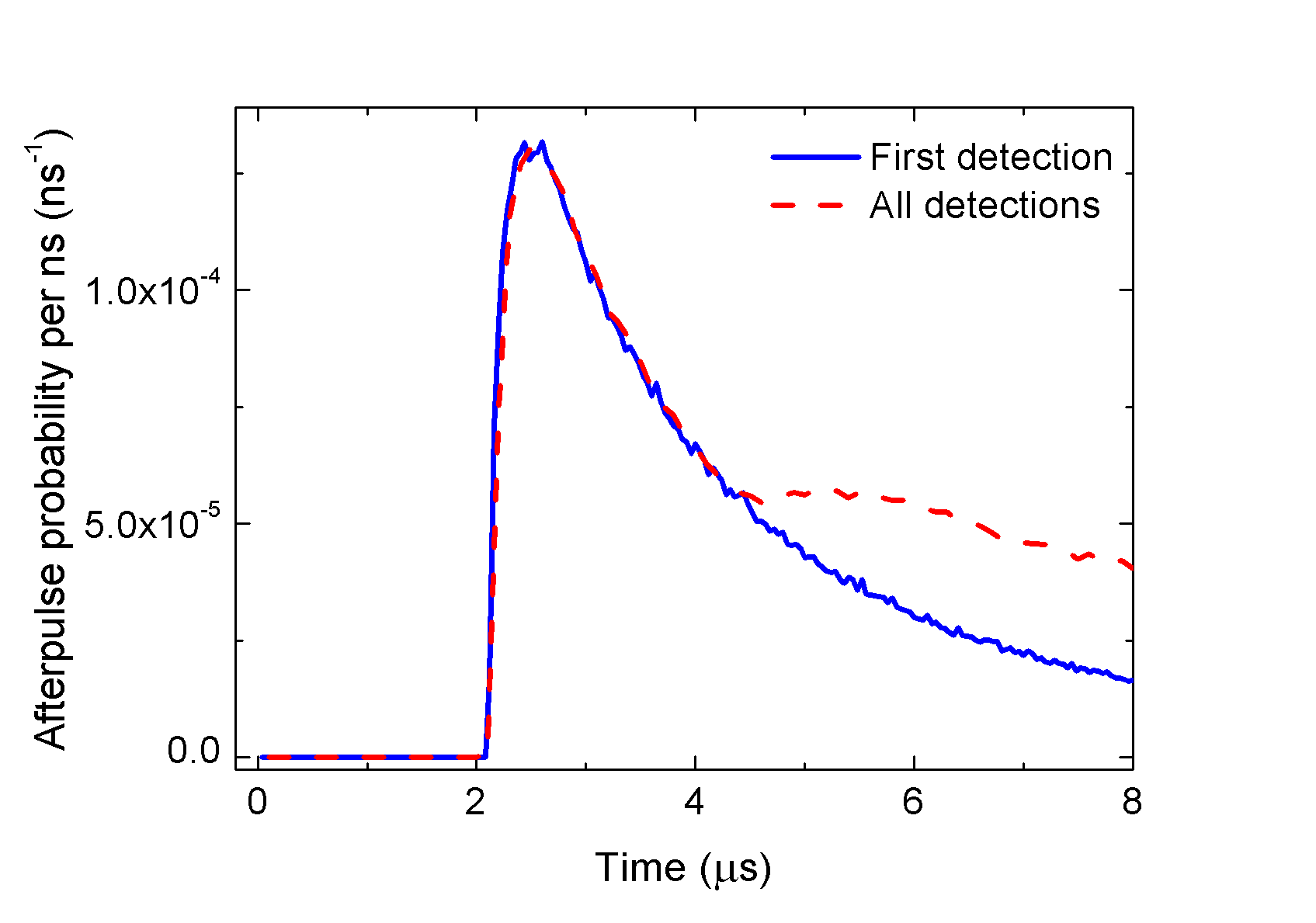}}\\
	\caption{(a) Experimental setup for the afterpulse characterization. The FPGA is  triggered by an external clock at 50 MHz. (b) Afterpulse acquisition obtained with the FPGA at 12\% of detection efficiency and $\tau_{HO}$ of 2 $\mu$s. The red dotted (blue solid) curve shows the afterpulse probability distribution considering (not considering) higher order afterpulses. }
\label{fig:3}
\end{figure}

Figure \ref{subfig:3:1} shows the measurement setup.  A 1550 nm laser diode (LD) triggered by the FPGA sends a 200 ps (FWHM) laser pulse to a 50/50 fibre beam splitter.  One output is used for monitoring the laser power in real-time. The other output goes to a variable attenuator which is set  in order to have on average 1 photon per pulse impinging on the photodiode. The electrical circuit is able to discriminate the avalanche signal and to apply a programmable hold-off time, $\tau_{HO}$.\\ 

Figure \ref{subfig:3:2} shows a typical histogram of the afterpulse probability vs time. In this measure a $\tau_{HO}$ of 2 $\mu$s was chosen.  Higher order afterpulses are significant: the blue solid line shows only the first afterpulses, the red dotted line includes all the afterpulses. After the laser detection no afterpulses can appear due to the hold off time of 2 $\mu$s. At 4 $\mu$s the second order effects start to show up. \\
Higher order afterpulses have to be taken into account in order to prevent overestimating of the detector efficiency. However, in order to characterize correctly the trap lifetimes, higher order afterpulses should be considered.\\ 

\begin{figure}[!htbp]
	\centering
	\includegraphics[width=11cm]{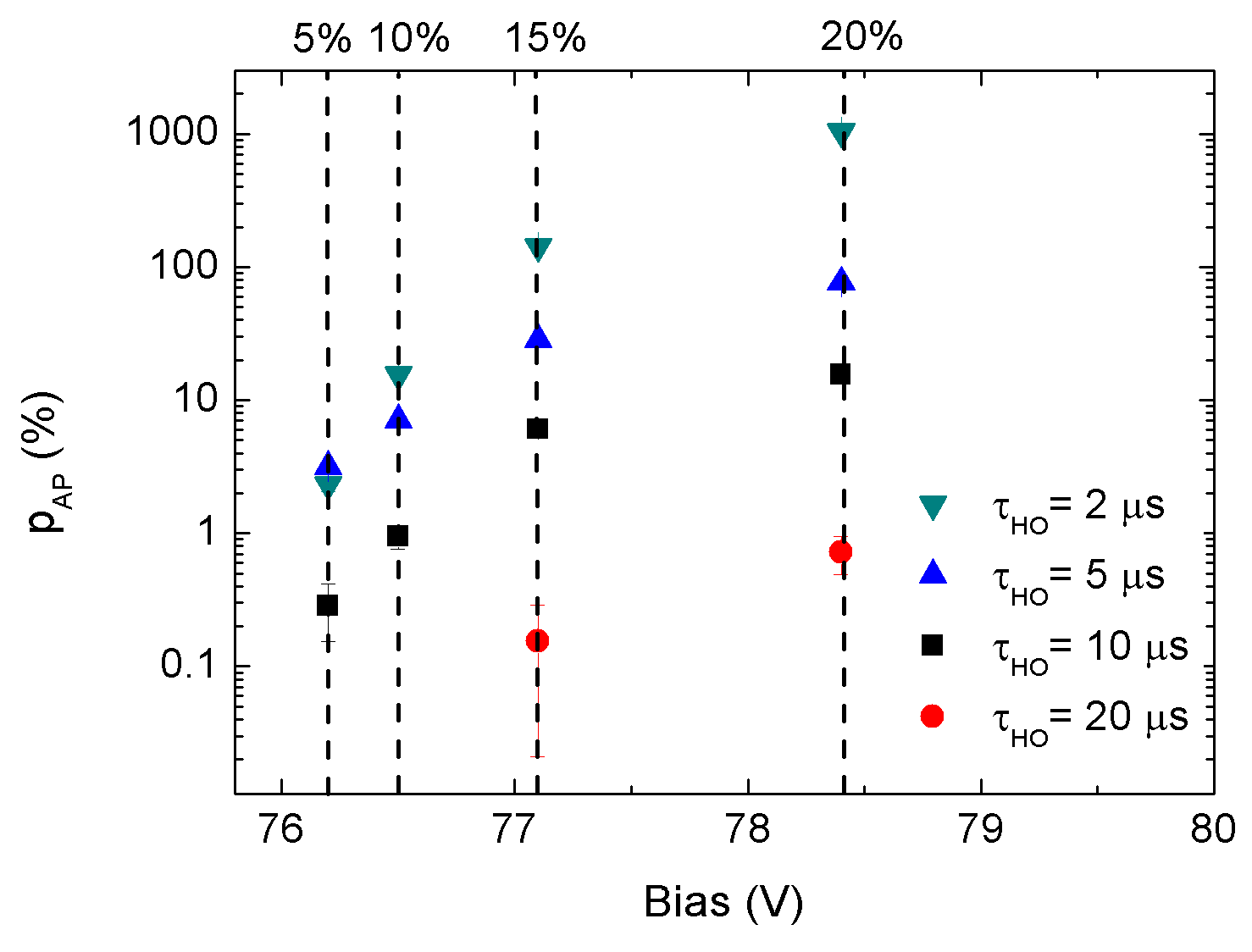}
	\caption{Total afterpulse probability (in \%) as a function of the bias voltage for different $\tau_{HO}$ time. At bias lower than 77 V for $\tau_{HO}$ of 20 $\mu$s the total afterpulse probability cannot be determined. For information, the approximate values of detection efficiency are indicated.}
	\label{fig:5}
\end{figure}

We investigated the total afterpulse probabilities (including the higher order afterpulses) by means of the FPGA for different biases and $\tau_{HO}$. In order to reduce the effect of afterpulses originating from earlier avalanches to a negligible level, we applied a \emph{waiting cycle} of 35 $\mu$s. Then we record the afterpulse distribution over a range of 35 $\mu$s after the detection (250 time-bins with a width of $\tau$=140 ns).  
The total afterpulse probability, p$_{AP}$, is derived by the formula:
\begin{equation}
p_{AP}=\sum_i\left(\frac{C_i}{C_d}-r_{dc}\cdot\tau\right)
\end{equation}
where $C_i$ is the number avalanches occurring in the time-bin $i$, $C_d$ is the number of laser detections and $r_{dc}$ is the dark count rate measured as described below. Figure \ref{fig:5} reports the total afterpulse probability as a function of the bias voltage for different $\tau_{HO}$. In the upper side of the graph the corresponding values of the detection efficiency have been reported(see below for further details).\\
Note that on the one hand, with $\tau_{HO}$ = 20 $\mu$s, the total afterpulse probability is below 1\% even at about  20$\%$ of detection efficiency. On the other hand,  p$_{AP}$ reaches 1000\% with $\tau_{HO}$ = 2 $\mu$s, i.e. each avalanche generates on average $\sim$10 afterpulses.\\

\section{Characterization of the free-running detector module}

\begin{figure}[!htbp]
	\centering
	\subfigure[]{\label{subfig:2:1} \includegraphics[width=7.5cm]{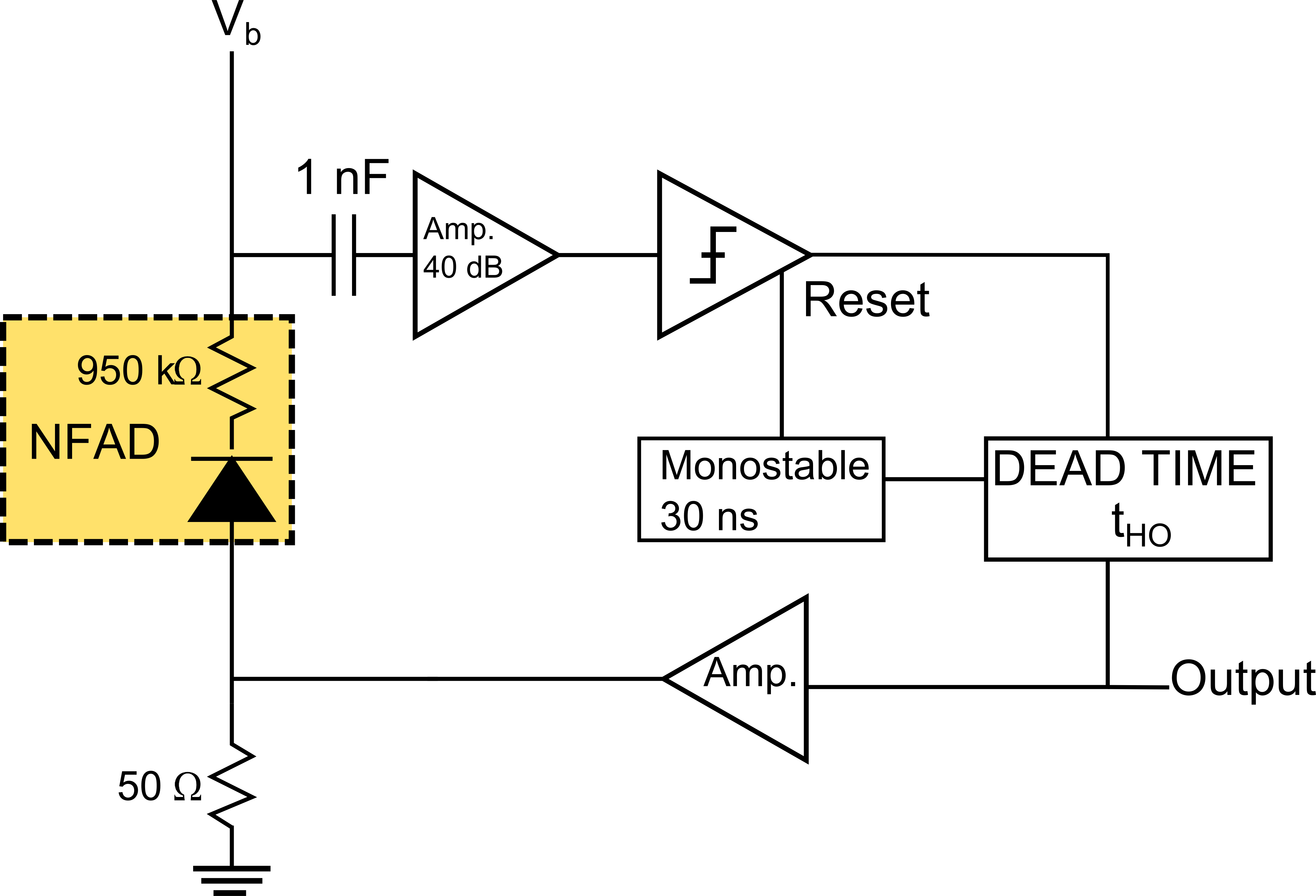}}\qquad
	\subfigure[]{ \label{subfig:2:3}\includegraphics[width=7.5cm]{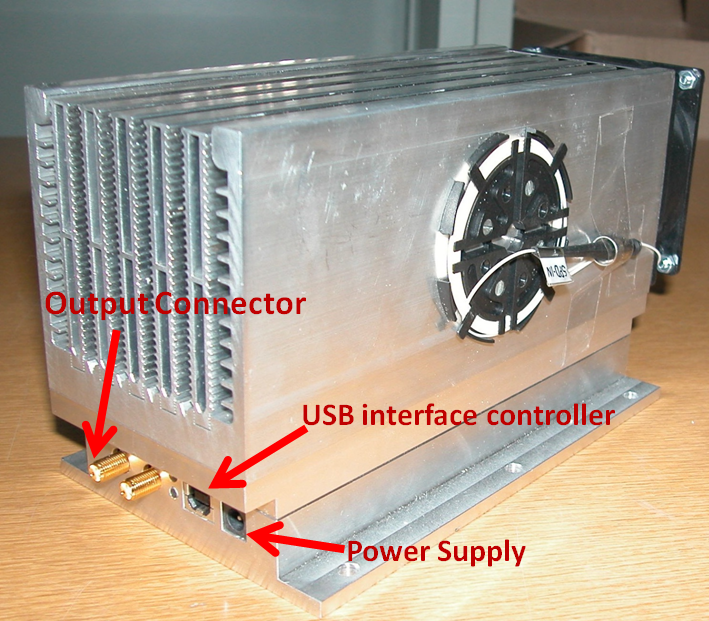}}\\
	\caption{(a) Electronic circuit of the detector module. (b)  Picture of the detector module.}
	\label{fig:2}
\end{figure}

The previous characterizations confirm the efficient quenching obtained by the monolithically integrated resistor. In the following we developed a versatile and simple detector module. We implemented a detector logic circuit, outlined in Fig.\ref{subfig:2:1}, able to discriminate the avalanches and apply the hold off. The diode cathode is capacitively coupled and the signal amplified by a 40 dB, large bandwidth (40 MHz - 6 GHz) amplifier. The voltage drop induced by the avalanche current is discriminated and digitalized providing a LVTTL signal as output. This signal is also used to apply a variable hold off with a duration that can be controlled. Due to the capacitive coupling, two spurious voltage spikes occur in correspondence with the rising and falling edges of the hold-off signal. While the first spike is positive and not discriminated the second spike, negative, could be discriminated. For this the discriminator is inhibited at the end of the hold-off signal. The diode is installed on a 4-stage thermoelectric cooler and together with the electronics sealed in a custom box shown in Fig.\ref{subfig:2:3}. The parameters (bias voltage, hold off, operating temperature etc.) can be set via the USB interface.\\
  
\begin{figure}[!htbp]
	\centering
	\subfigure[]{\label{subfig:4:1} \includegraphics[width=7.5cm]{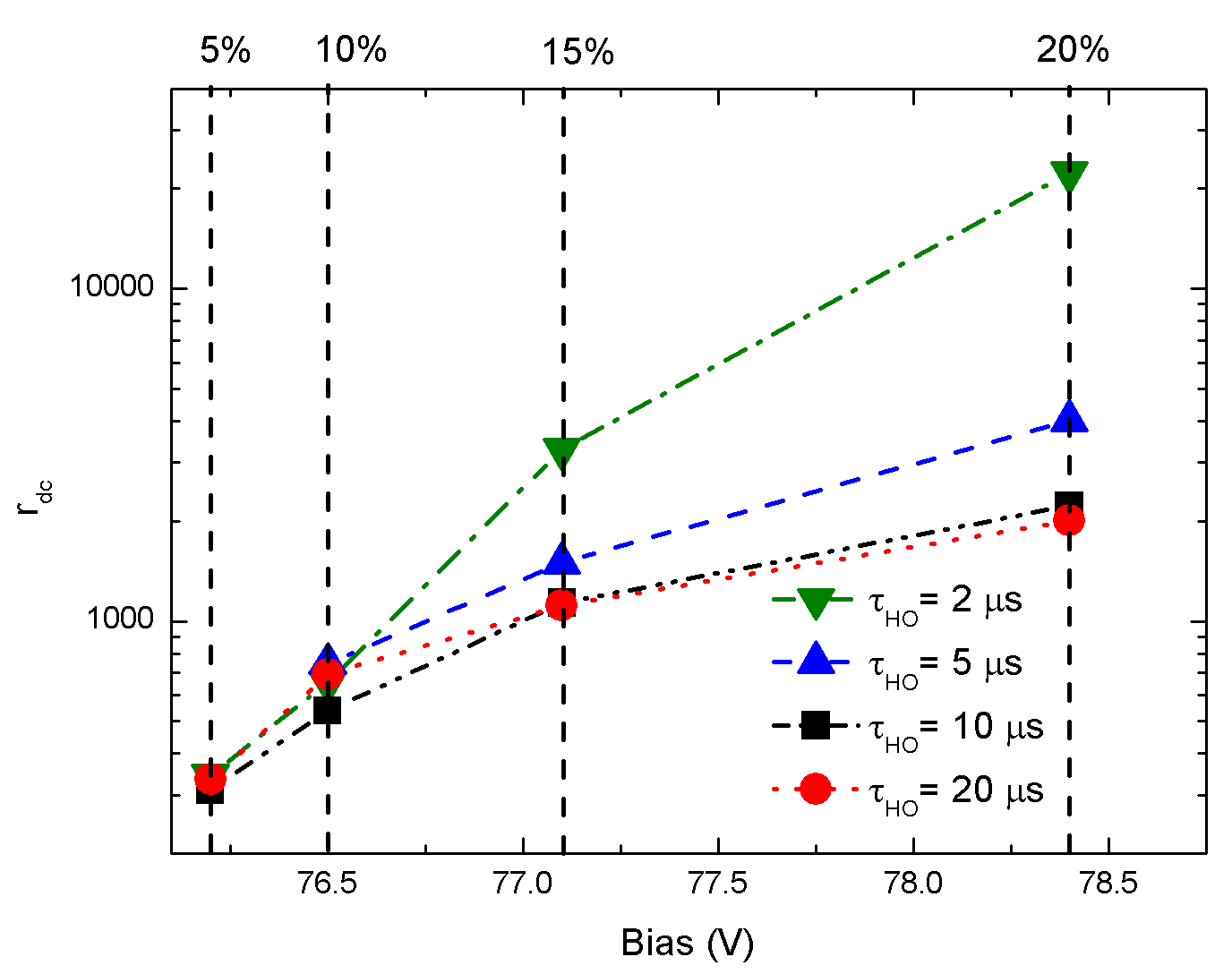}}\qquad
	\subfigure[]{ \label{subfig:4:3}\includegraphics[width=7.5cm]{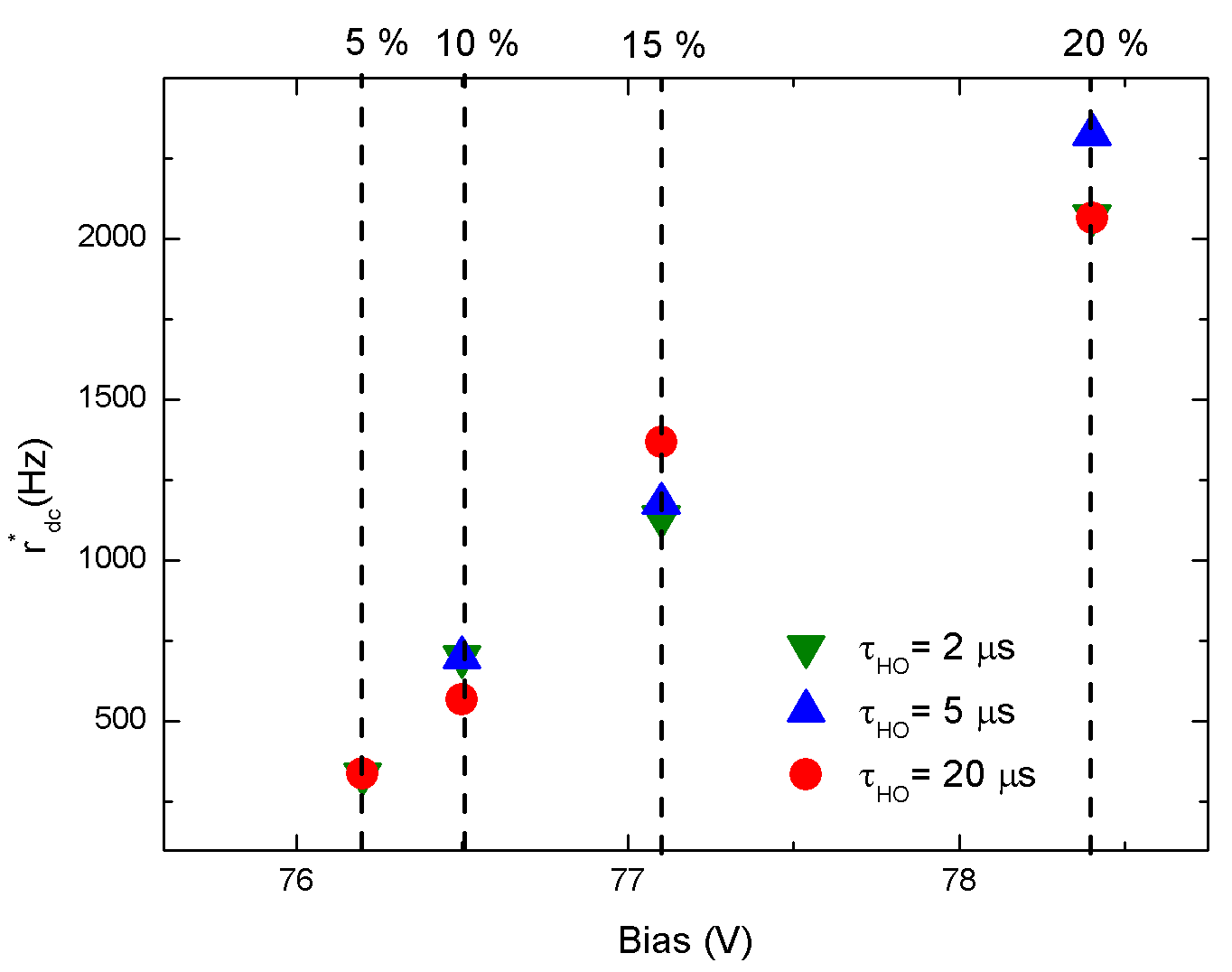}}\\
	\caption{(a) Measured dark-count rate, $r_{dc}$, as a function of the voltage bias for different hold-off times, $\tau_{HO}$. The graph shows the effectiveness of the negative feedback resistor with $r_{dc}$ yielding 600 Hz at 10\% of detection efficiency, $\eta$. At higher bias voltages, the dark-count rate increases reducing the hold-off time. This phenomenon has been attributed to the afterpulses. (b) The inherent dark-count rate (corrected for afterpulsing effect and hold-off time following Eq.\ref{eq:3}), $r^*_{dc}$, as a function of the bias voltage for hold off times yielding 5 $\mu$s and 2 $\mu$s and 20 $\mu$s. As expected, $r^*_{dc}$ does not change with the hold-off time. This confirms Eq.\ref{eq:3} demonstrating that the differences between the measured dark-count rate shown in Fig.\ref{subfig:4:3} are mainly due to afterpulses.}
	\label{fig:4}
\end{figure}

Now, we characterize the detector performances, namely the dark count rate and detection efficiency. The setup is very similar to the one shown in Fig. \ref{subfig:3:1}, with a continuous laser and no FPGA. The output of the module is simply connected to a counter. The characterization of a free-running InGaAs photon counter is not straightforward, since one has to take into account the afterpulsing and apply a correction for the hold-off time. As we will show the detection efficiency is easily overestimated if it is not correctly compensated for afterpulsing. Moreover, due to rather high hold-off time the detector saturates quickly and the effective efficiency is a function of the count rate. The same is true for the afterpulsing fraction, which decreases with higher count rates.
In the following referring to the detection efficiency, $\eta$, we mean the probability that a photon, already coupled inside the optical fibre of the pigtailed detector, will originate a click, given that the detector is ready to detect. \\ 
First, we measure the $r_{dc}$ (with closed shutter) as a function of bias voltage for different $\tau_{HO}$ (see Fig. \ref{subfig:4:1}). We note a dramatic increase of dark count rate for short $\tau_{HO}$ and high bias voltage. This is the clear signature of afterpulsing. However, the inherent dark-count rate, $r^*_{dc}$, due to thermal generation of carriers or tunnelling should obviously be independent of $\tau_{HO}$. We can check this by correcting the measured $r_{dc}$ for the $p_{AP}$ measured before and for the hold-off time (sometimes called \textit{dead-time correction}). The inherent dark count rate is calculated as follows:\\

\begin{equation}
r^*_{dc}=\frac{r_{dc}}{(1+p_{AP})\cdot(1-r_{dc}\cdot \tau_{HO})}
\label{eq:3}
\end{equation}\\

We compared these results with the dark count rate measured at 20$\mu$s of $\tau_{HO}$ time where the afterpulses are always negligible. Figure \ref{subfig:4:3} shows that this correction makes the $r^*_{dc}$ almost independent from the $\tau_{HO}$, confirming that the measured $p_{AP}$ are correct.\\

As we said before the number of clicks, $S$, originated by the detector depends not only by the detection efficiency but also by afterpulsing and the hold-off time $\tau_{HO}$.\\
Basing on this, $S$ is given by the following formula:
\begin{equation}
S=(n\cdot\eta+r^*_{dc})(1-S\cdot \tau_d)(1+p_{AP})
\end{equation}
where, $n$ is the number of photons per second, yielding 10$^5$ photons/s in our case. $\eta$ can thus be easily derived from:
\begin{equation}
\eta=\frac{1}{n} \left( \frac{S}{(1-S \cdot \tau_d) \cdot (1+p_{AP})}  - r^*_{dc} \right)
\label{eq:1}
\end{equation}
This formula allows to correct the extra counts originated by the afterpulse and take into account the hold-off time corrections at the same time. 

\begin{figure}[!htbp]
	\centering
	\subfigure[]{\label{subfig:7:1} \includegraphics[width=7.5cm]{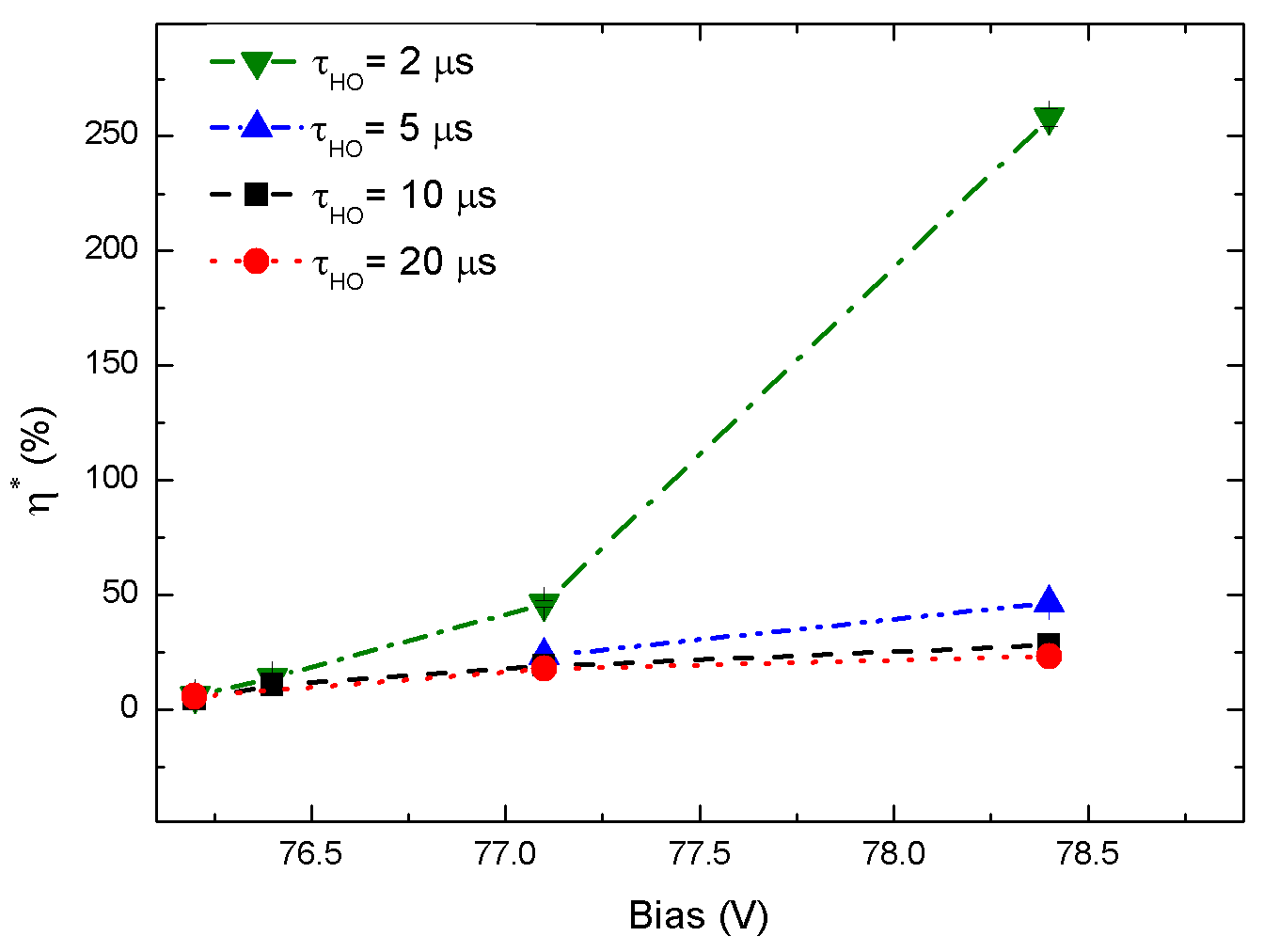}}\qquad
	\subfigure[]{ \label{subfig:7:2}\includegraphics[width=7.5cm]{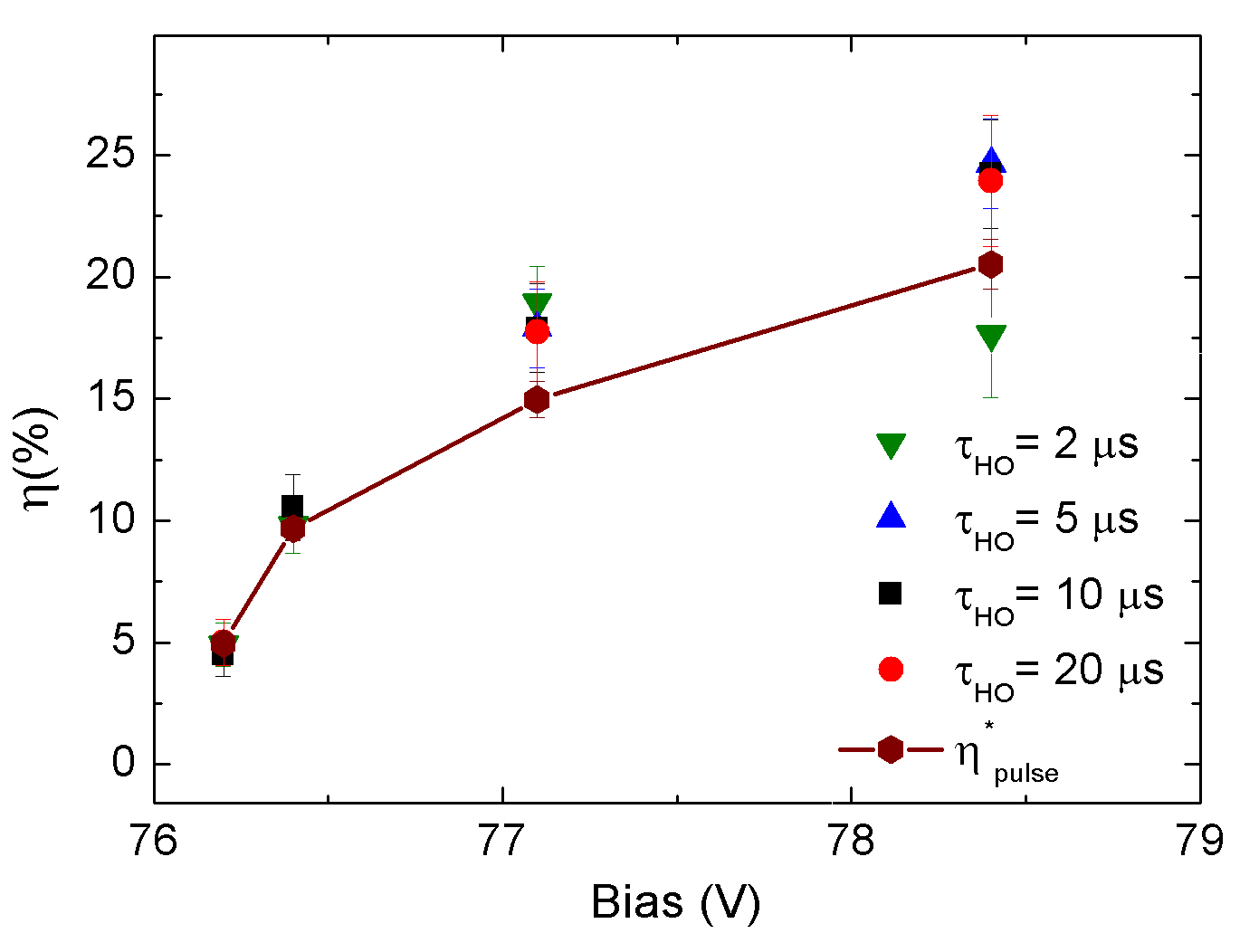}}
	\caption{Detection efficiencies for four different hold-off times. (a) The detection efficiency $\eta'$ corrected only for the hold-off time $\tau_{HO}$(according Eq.\ref{eq:1} for $p_{AP}$ = 0 \cite{Thew}). As can be seen, increasing the bias voltage, the afterpulses increase leading to an overestimation of the detection efficiency. (b) The detection efficiency $\eta$, including also the afterpulse correction. Note that the dependency from the hold-off time $\tau_{HO}$ is drastically reduced.  As a comparison detection efficiency evaluated with the standard synchronous measurement for the gated mode by means of the FPGA is plotted as well (dark red hexagons-solid lines curve). The good agreement demonstrates that the measured afterpulse probabilities are correct.}	
\label{fig:7}
\end{figure}

As in the case of the $r_{dc}$ the afterpulsing correction becomes crucial in order to provide a correct evaluation for the detection efficiency. This can be immediately verify in Fig.\ref{subfig:7:1}: assuming $p_{AP}=0$ the detection efficiency (named $\eta'$ in order to distinguish it from $\eta$) becomes unrealistic for higher bias voltages and low $\tau_{HO}$. This dependence on $\tau_{HO}$ disappears considering $\eta$ (see Fig .\ref{subfig:7:2}).

\newpage

We can compare the obtained detection efficiencies in free-running mode with the detection efficiency obtained with the standard synchronous measurement for the gated mode. Obviously, they should be identical. We can mimic the gated mode operation with the setup we used for the afterpulsing characterization. We calculate the detection efficiency with the standard formula \cite{Jun2} correcting for dark counts and Poisson distribution of the photon number in the laser pulse:
\begin{equation}
\eta_{gated}=\frac{1}{\mu}\ln\left({\frac{1-\frac{r_{dc}}{Clk}}{1-\frac{C_d}{C_{lp}}}}\right)
\label{eq:2}
\end{equation}

where $\mu$ is the average number of photon per pulse (1 in our experiment), $r_{dc}$ is the dark count rate, $Clk$ is the clock frequency, $C_d$  is the number of detections coincident with the pulse sending and $C_{lp}$ is the number laser pulse triggered by the FPGA. We can see in Fig. \ref{subfig:7:1} that this measurement is consistent with the measurement in free-running mode. 

\begin{figure}[!htbp]
	\centering
	\includegraphics[width=10cm]{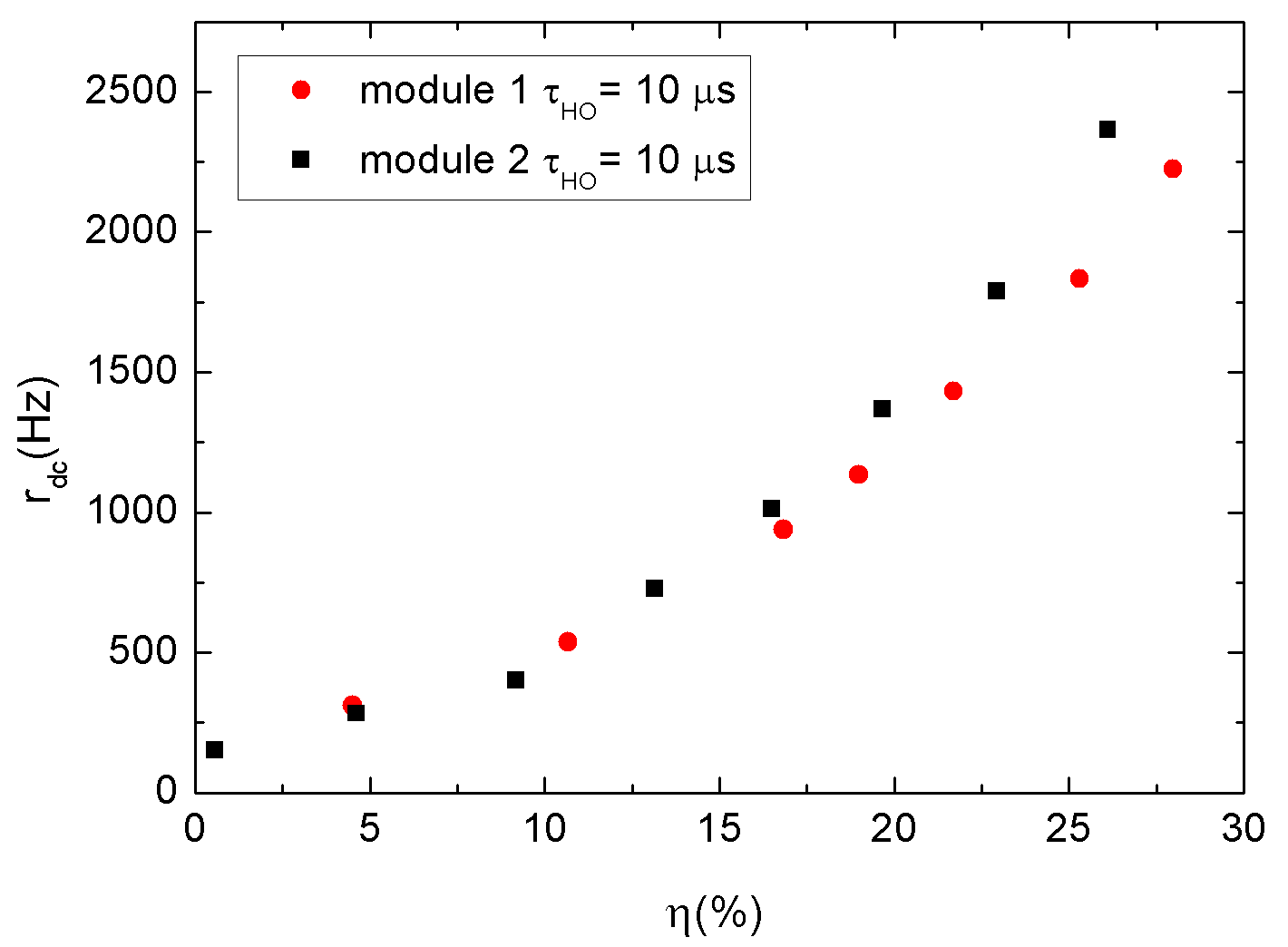}
	\caption{$r_{dc}$ as a function of the detection efficiency for two different modules (red circles and black squares) applying $\tau_{HO}$ = 10 $\mu$s. At 10\%  of detection efficiency the $r_{dc}$ yield about 600 Hz corresponding to 6$\cdot 10^{-7}$ counts/ns.}
	\label{fig:8}
\end{figure}

Finally, we plot the dark-count rate as a function of the detection efficiency for two modules with different APD from the same wafer. We can see very similar performances, see Fig.\ref{fig:8}. The effectiveness of the negative feedback resistor reduces the extra counts originated by the afterpulse allowing the light detection with reduced noise: at 10\% of $\eta$ the $r_{dc}$ yields only $\sim$ 600 Hz (corresponding to a dark count rate per ns of 6$\cdot$10$^{-7}$ counts/ns) and it increases up to $\sim$ 1500 Hz (1.5$\cdot$ 10$^{-6}$ counts/ns) at 20\% of $\eta$ both with $\tau_{HO}$ = 10 $\mu$s . Figure \ref{fig:8} reports the dark count rate $r_{dc}$ instead of the inherent dark count rate $r^*_{dc}$ since we believe the former more relevant from the practical point of view, representing the raw number of clicks outcoming from the detector when no light is sent without any further, mathematical, manipulations.\\ 
The $r_{dc}$ of the NFAD detector in free-running condition can be compared with that of the other detectors recently published in literature \cite{Jun2, Thew, Namekata, Dixon}. The comparison between the $r_{dc}$ at the same detection efficiency, demonstrates our detector highly competitive with the other considering also the  advantage of a simpler electronics. We would like also to remark that only the detector described in \cite{Thew}, a standard diode actively quenched, can be used in free-running condition with $r_{dc}$ yielding 1.2$\cdot 10^{-5}$ counts/ns  at the same efficiency (10\%) and hold-off time. This large difference is almost removed with longer hold-off time demonstrating that the improvement is only due to the effectively afterpulse suppression.\\
The NFAD performance can also be compared with that of a standard SNSPD: at the best signal-to-noise conditions the $r_{dc}$ for the superconducting detectors usually yields about 10$^-8$ counts/ns, but the detection efficiency rarely overcomes the 10\% especially in the commercial devices. Moreover, the increasing of the bias generally produces an exponential increase of the noise and the dark-count rate rapidly surpasses the NFAD one at the same efficiency, e.g. in \cite{Miki} the  $r_{dc}$ yields 10$^{-5}$ counts/ns at 15\% of detection efficiency.\\

\newpage

\section{Timing Resolution}
\begin{figure}[htbp]
	\centering
	\includegraphics[width=10cm]{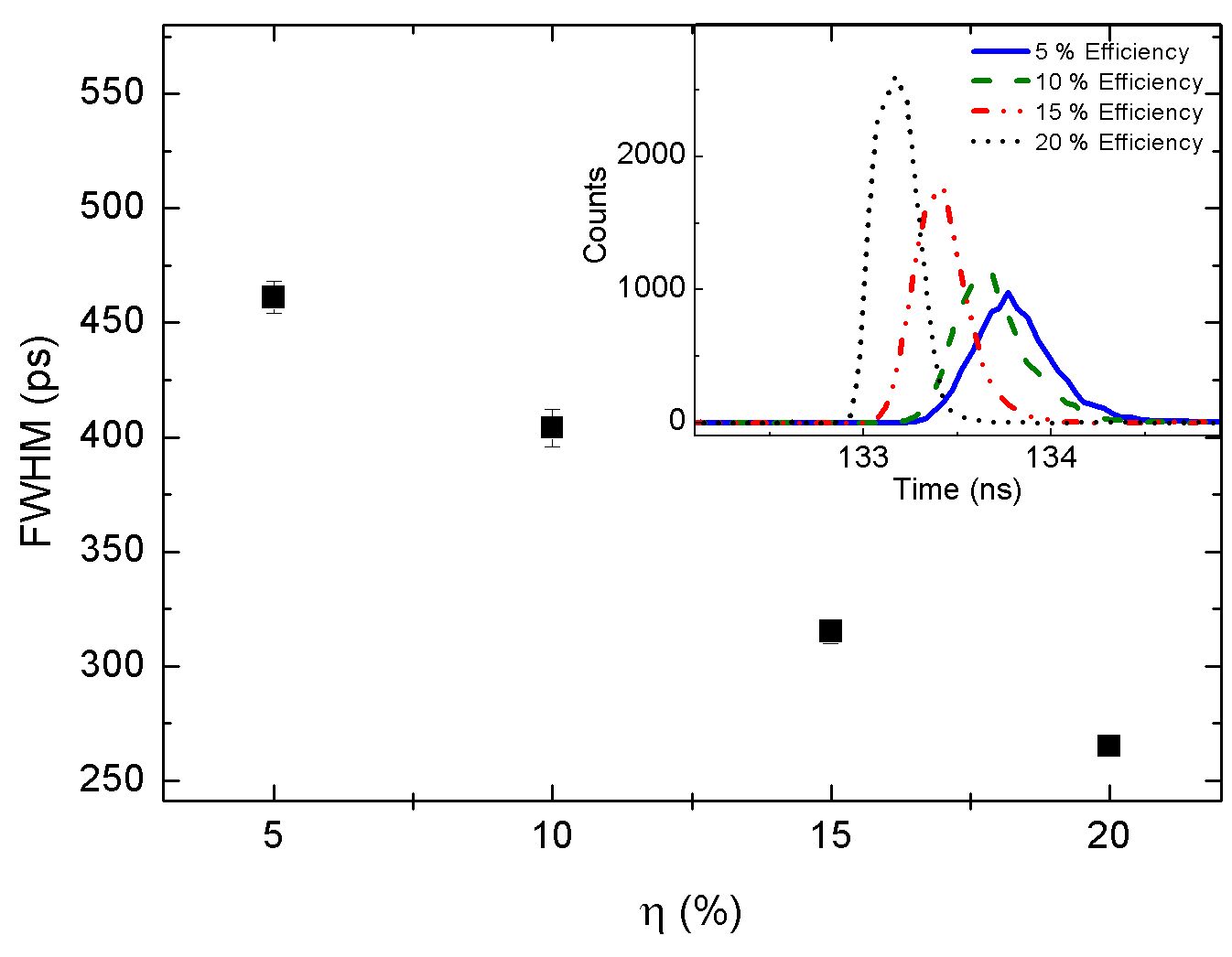}
	\caption{Jitter as a function of the $\eta$. The timing jitters were expressed by the Full Width at Half Maximum, FWHM of the TDC acquisition. In the inset was reported the time measurements performed with the TDC at the different $\eta$.}
	\label{fig:9}
\end{figure}

The timing resolution of the detector was measured using a ps laser and a Time to Digital Converter, TDC. The laser pulses have a length of about 33 ps. We work at 0.1 photons on average, at a pulse rate of 10 KHz and the APD is at 223 K. We record the histogram of the time delays between the laser trigger and the detection signal from the detector module for different $\eta$ (see inset in Fig.\ref{fig:9}).

The timing resolution or jitter is in good approximation just the FWHM of the histogram, as the pump pulse width can be neglected. 
We observed a jitter decreasing with increasing the $\eta$. We believe that higher bias voltage helps the avalanches to reach the detection threshold with less temporal variation. The jitter values comprise between 250 ps and 450 ps for $\eta$  between 5\% and 20\%,  respectively. Those results are consistent with the value provided by Princeton Lightwave demonstrating that the jitter introduced by our electronics is negligible respect to the diode. 

\section{Conclusion}

In conclusion, we developed and characterized a compact, free running detector module  based on a NFAD detector.  We also improved the standard double-window technique for the afterpulsing characterization. Our algorithm implemented by a FPGA allows to put the APD in a well-defined initial condition and to measure the impact of the higher order afterpulses. The user can easily adapt the  bias voltage and hold-off time in order to obtain an optimum performance for his application. For instance, at 10\% quantum efficiency, a dark count rate as low as $\sim$ 600 Hz is achieved with a hold-off time of only 10 microseconds. This makes this module a cheap and convenient alternative to SNSPD detectors. This module is now commercially available\cite{IDquantique}. Finalizing this manuscript we became aware of similar work \cite{Yuan}.

\section*{Acknowledgements}
We thank S. Decadri for the work done during the electronic development. This work was
supported by the Swiss National Science Foundation NCCR within the projects Quantum Photonics
and QSIT, Quantum Science and Information Technology.

\end{document}